\documentclass[11pt]{article}

\input epsf
\usepackage{latexsym}
\usepackage{amssymb}
\usepackage{amsmath}
\usepackage[dvips]{graphicx}
\usepackage{array}

\long\def\symbolfootnote[#1]#2{\begingroup%
\def\thefootnote{\fnsymbol{footnote}}\footnote[#1]{#2}\endgroup} 

\begin{document}
\noindent KCL-PH-TH/2011-{\bf 20}

\vskip 0.5 cm 

\centerline{\Large \bf Effects of anisotropic dynamics on cosmic
  strings}
                                                                              
\vskip 1cm

\centerline{Kerstin E. Kunze
\symbolfootnote[1]{\tt kkunze@usal.es}}
\vskip 0.3cm \centerline{{\sl Departamento de F\'\i sica Fundamental
    and IUFFyM, Universidad de Salamanca,}} \centerline{{\sl Plaza de
    la Merced s/n, 37008 Salamanca, Spain }}

\vskip 0.5cm

\centerline{{and}}

\vskip 0.5cm

\centerline{Mairi Sakellariadou
\symbolfootnote[2]{\tt mairi.sakellariadou@kcl.ac.uk}}
\vskip 0.3cm \centerline{{\sl Department of Physics, King's College
    London, University of London}} \centerline{{\sl Strand WC2R 2LS,
    London, U.K.}}

\vskip 1.5cm

%%%%%%%%%%%%%%%%%%%%%%%%%%%%%%%%%%%%%%%%%%%%%%%%%%%%%%%%%%%%%%%%%%%%%%
\centerline{\bf Abstract}
\vskip 0.5cm
\noindent
The dynamics of cosmic strings is considered in anisotropic
backgrounds. In particular, the behaviour of infinitely long straight
cosmic strings and of cosmic string loops is determined.  Small
perturbations of a straight cosmic string are calculated. The
relevance of these results is discussed with respect to the possible
observational imprints of an anisotropic phase on the behaviour of a
cosmic string network.

\vskip 1cm
%%%%%%%%%%%%%%%%%%%%%%%%%%%%%%%%%%%%%%%%%%%%%%%%%%%%%%%%%%%%%%%%%%%%%%

\section{Introduction}
\setcounter{equation}{0}

Topological defects are expected to arise naturally in the early
universe, as the result of phase transitions followed by spontaneously
broken symmetries. In the context of gauge theories, monopoles and
domain walls may lead to disastrous consequences for cosmology, while
cosmic strings may play a useful r\^ole. Cosmic strings are linear
topological defects, analogous to flux tubes in type-II
superconductors, or to vortex filaments in superfluid helium. Cosmic
strings were shown \cite{Jeannerot:2003qv} to be generically formed at
the end of an inflationary era, within the framework of supersymmetric
grand unified theories.

Observations indicate that the universe is nearly homogeneous and
isotropic on large scales. This can naturally be achieved by a stage
of de Sitter inflation at early times.  In the presence of a positive
cosmological constant the isotropic solution is an attractor solution,
as was shown in the case of spatially homogeneous models which are
described by the Bianchi models. All but Bianchi IX models approach
the de Sitter solution at late times \cite{wald}. The study of
observational imprints, such as the primordial curvature perturbation
spectrum, has largely focused on the isotropic phase. However,
recently the effects of an anisotropic stage on the anisotropies of
the cosmic microwave background have been studied
\cite{gkp}-\cite{wks}.  One of the motivations was the observation of
anomalies in the WMAP data, such as the alignment of the moments in
the lowest multipoles and the suppression of the quadrupole
\cite{gkp}.  In Ref.~\cite{jaffe} the WMAP first year data were fitted
to a Bianchi VII$_{h}$ model and thus accounting for the
aforementioned anomalies by assuming a globally anisotropic
background.  The temperature anisotropies of the cosmic microwave
background in spatially homogeneous and anisotropic universes were
first studied in Ref.~\cite{barr}.

In the context of topological defects, the effects of an anisotropic
stage of inflation on domain walls was studied in Ref.~\cite{am1} and
on cosmic string loops in Ref.~\cite{am2}.  
In Ref. \cite{as} the equations of motion for cosmic strings in a (3+1)-dimensional
Friedmann-Lema\^{i}tre-Robertson-Walker model with extra dimensions were presented
which describe in general an anisotropic higher dimensional space-time.
There the evolution of a cosmic string network was studied in a $(D+1)$-dimensional isotropic space-time
as well as in a (3+1)-dimensional  Friedmann-Lema\^{i}tre-Robertson-Walker model with static 
extra dimensions.
In what follows we investigate the fingerprint of
anisotropic expansion on cosmic strings dynamics in four
dimensions. Such a study may be relevant for models where the duration
of inflation did not have enough time to reach the attractor solution,
so that cosmic strings, formed at the end of inflation, live in an
anisotropic background.

%%%%%%%%%%%%%%%%%%%%%%%%%%%%%%%%%%%%%%%%%%%%%%%%%%%%%%%%%%%%%%%%%%%%%%

\section{Strings in anisotropic backgrounds}

The simplest spatially homogeneous, anisotropic model is of type
Bianchi I which is described by the metric
\begin{eqnarray}
ds^2=dt^2-a_1^2dx^2-a_2^2dy^2-a_3^2dz^2~,
\end{eqnarray}
where $a_i(t)$ with $i=1,2,3$ are the scale factors in the three
different spatial directions.  Inflationary solutions in this
background coupled to a scalar field $\varphi$ with a potential
$V(\varphi)$ were studied in detail in Ref.~\cite{pcu2}. In
particular, in the presence of a positive cosmological constant,
corresponding to $V={\rm const.},$ $\dot{\varphi}=0$, the solutions
for the scale factors $a_i$ are given by \cite{pcu2}
\begin{eqnarray}
a_i(t)=A\left[\sinh\left(\frac{t}{t_*}\right)\right]^{\frac{1}{3}}
\left[\tanh\left(\frac{t}{2t_*}\right)\right]^{(2/3)\sin
  \beta_i}~,
\end{eqnarray}
where $A\equiv({\cal K}^2/6V_0)^{\frac{1}{6}}$, where ${\cal K}^2$ is
an integration constant related to the square of the shear, $V_0\equiv
8\pi V/(3M_{\rm P}^2)$ with $M_{\rm P}$ the Planck mass, and
$t_*\equiv 1/3\sqrt{V_0}$.  Moreover, $\beta_i=\beta+(2\pi/3)i$,
$i\in\{1,2,3\}$ with $\beta$ a parameter, which we choose, following
Ref.~\cite{pcu2}, to be between 0 and $2\pi/3$. 

\begin{figure}[ht]
\centerline{\epsfxsize=2.8in\epsfbox{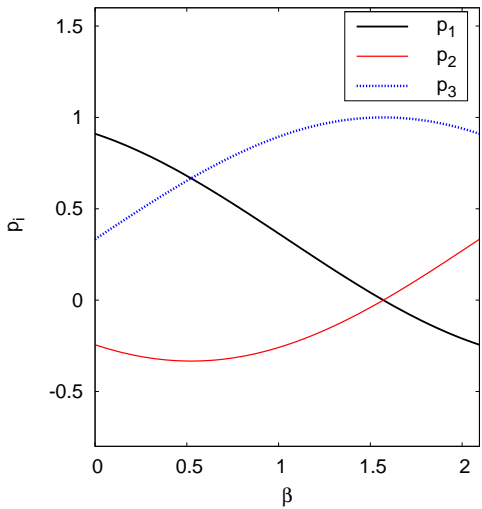}}
\caption{The Kasner exponents as a function of $\beta$.}
\label{fig0}
\end{figure}

In these backgrounds there is an initial singularity and the behaviour
of the scale factors close to the singularity is described by a Kasner
solution \cite{pcu2}
\begin{eqnarray}
a_i(t)=a_{i,*}\left(\frac{t}{2t_*}\right)^{p_i}~,
\end{eqnarray}
where in the following $t_*$ will be rescaled so to absorb the factor 2.
The exponents $p_i$ satisfy the Kasner relations
\begin{eqnarray}
\sum_i p_i=1
\hspace{1cm}
,
\hspace{1cm}
\sum_i p_i^2=1~.
\end{eqnarray}
The Kasner exponents are given by $p_i=(2/3)\sin\beta_i+(1/3)$ and are
shown \cite{pcu2} in Fig.~\ref{fig0}.

In the following we will use the metric in the form
\begin{eqnarray}
ds^2=a_1^2(\tau)[d\tau^2-dx^2]-a_2^2(\tau)dy^2-a_3^2(\tau)dz^2~,
\end{eqnarray}
where in the case of an isotropic space-time $\tau$ corresponds to
conformal time.  Transforming these solutions to the time variable
$\tau$, the scale factors take the form
\begin{eqnarray}
a_i(\tau)=a_{i,*}\left(\frac{\tau}{\tau_*}\right)^{\alpha_i}~,
\label{scale}
\end{eqnarray}
where $\alpha_i\equiv p_i/(1-p_1)$.

It should be noted that $p_1<1$ (cf., Fig.~\ref{fig0}) and therefore
the expanding or contracting nature, respectively, of a direction only
depends on the sign of $p_i$ and does not change when changing from
cosmic time to the conformal time-like variable $\tau$. Moreover, it
is interesting to note that the direction whose dynamics is determined
by the Kasner exponent $p_3$ is always expanding.  Whereas the
direction associated with $p_1$ changes its behaviour from an
expanding to a contracting universe, the direction determined by $p_2$
changes from a contracting to an expanding universe. The cross-over
takes place for both directions at $\beta=\pi/2$, in which case these
directions are actually static and $p_3$ reaches its maximum at
$p_3=1$.

Our aim is to estimate the effects of an anisotropic stage on the
dynamics of a cosmic string. Therefore, for simplicity, the Kasner
solution is used in Sections \ref{sec3} and \ref{sec4} to describe the
background dynamics.  Moreover, in the numerical solutions in Section
\ref{sec3} we set $\tau_*=1$ and $a_{i,*}=1$.  The discussion of
string loops in Section \ref{sec5} is done for general anisotropic
space-times.

Neglecting the thickness of the string, the dynamics of the string is
determined by the Nambu-Goto action (cf., e.g., Ref.~\cite{vs})
\begin{eqnarray}
S=-\mu\int \sqrt{-\gamma} d^2\zeta~,
\label{ng}
\end{eqnarray}
where $\mu$ is the string mass per unit length and $\gamma$ is the
determinant of the worldsheet metric.
Varying the action, Eq.~(\ref{ng}), w.r.t. the space-time coordinates
$x^{\mu}(\zeta^a)$ gives the string equation of motion.  Usually the
following gauge conditions are imposed
\begin{eqnarray}
\zeta^0\equiv\tau\hspace{1cm},\hspace{1cm}
\dot{x}\cdot x'=g_{\mu\nu}\dot{x}^{\mu}x'^{\;\nu}=0~,
\label{gauge}
\end{eqnarray}
where a dot denotes $\partial/\partial\tau$ and a prime
$\partial/\partial\zeta$ with $\zeta\equiv\zeta^1$.  Then the string
equation of motion is given by \cite{vs}
\begin{eqnarray}
\frac{\partial}{\partial\tau}\left[\frac{\dot{x}^{\mu}x'^{\;2}}{\sqrt{-\gamma}}
\right]+\frac{\partial}{\partial\zeta}
\left[\frac{x'^{\;\mu}\dot{x}^2}{\sqrt{-\gamma}}
\right]+\frac{1}{\sqrt{-\gamma}}\Gamma^{\mu}_{\nu\sigma}
\left(x'^{\;2}\dot{x}^{\nu}\dot{x}^{\sigma}
+\dot{x}^2x'^{\;\nu}x'^{\;\sigma}\right)=0~.
\label{sd1}
\end{eqnarray}
The non-vanishing components of the world-sheet metric
$\gamma_{ab}=g_{\mu\nu}x^{\mu}_{\;\; ,a} x^{\nu}_{\;\; ,b}$ (latin
indices run from 1 to 3) are given by
\begin{eqnarray}
\gamma_{\tau\tau}&=&\dot{x}^2=a_1^2-\sum_{i}a_i^2\dot{x}^{i\;2}~,\nonumber\\
\gamma_{\zeta\zeta}&=&x'^{\; 2}=-\sum_i a_i^2x'^{i\; 2}~.
\end{eqnarray}
Defining
\begin{eqnarray}
\epsilon\equiv-\frac{x'^{\;2}}{\sqrt{-\gamma}}~,
\label{eps}
\end{eqnarray}
Eq.~(\ref{sd1}) yields 
\begin{eqnarray}
\dot{\epsilon}=-\epsilon\left[\frac{\dot{a_1}}{a_1}+\sum_i
\frac{a_i\dot{a}_i}{a_1^2}
\left(
\dot{x}^{i\;2}
-\left(\frac{x'^{\;i }}{\epsilon}\right)^2
\right)
\right]~,
\label{e1}
\end{eqnarray}
for $\mu=0$ and
\begin{eqnarray}
\ddot{x}^i+\left[2\frac{\dot{a}_i}{a_i}+\frac{\dot{\epsilon}}{\epsilon}
\right]
\dot{x}^i=\frac{1}{\epsilon}\left(\frac{x'^{\;i}}{\epsilon}\right)'~,
\label{v}
\end{eqnarray}
for $\mu=i$.  In the isotropic limit the usual equations of a cosmic
string in a flat Friedmann-Lema\^{i}tre-Robertson-Walker background
are recovered (cf, .e.g., Ref.~\cite{vs}). Moreover, these equations agree
with those  in Refs.~\cite{am2} and \cite{as}.

%%%%%%%%%%%%%%%%%%%%%%%%%%%%%%%%%%%%%%%%%%%%%%%%%%%%%%%%%%%%%%%%%%%%%%

\section{A moving straight string}
\label{sec3}

One can show that here, as for an isotropic background, a static
straight string $x^i(\zeta)=c^i\zeta$ is a solution (cf., e.g.,
Ref.~\cite{vs}).  Following Ref.~\cite{vs}, let us consider a moving
straight string of the form
\begin{eqnarray}
x^i(\zeta,\tau)=b^i(\tau)+c^i\zeta~,
\end{eqnarray}
where $c^i$ are constants.  Defining
$v^i=\dot{b}^i$, Eq.~(\ref{v}) yields
\begin{eqnarray}
\dot{v}^i+\left[2\frac{\dot{a}_i}{a_i}-\frac{\dot{a}_1}{a_1}
  -\sum_j\frac{a_j\dot{a}_j}{a_1^2}\left(v^{j\;\;2}
  -\left(\frac{c^i}{\epsilon}\right)^2\right)\right]v^i=0~.
\label{v1}
\end{eqnarray}
In an isotropic background with a scale factor $a(t)$ the variable $v^i$ corresponds to the proper velocity of the
string $\vec{V}= a\frac{d\vec{x}}{dt}=\frac{d\vec{x}}{d\tau}=\vec{v}$. However, in an anisotropic background the physical 
velocity of the string is given by $V^i=a_i\frac{dx^i}{dt}=\frac{a_i}{a_1}v^i$.
From the definition of $\epsilon$ (cf., Eq.~(\ref{eps})) it is found
that
\begin{eqnarray}
\epsilon=\left(
\frac{\sum_i a_i^2(c^i)^2}{a_1^2-\sum_i a_i^2(v^i)^2}
\right)^{\frac{1}{2}}~,
\label{eps1}
\end{eqnarray}
so that $\epsilon=\epsilon(\tau)$.  In the isotropic case
Eq.~(\ref{v1}) can be solved to find the behaviour of $v=|\vec{v}|$,
namely \cite{vs}
\begin{eqnarray}
v(1-v^2)^{-\frac{1}{2}}\propto a^{-2}~.
\label{iso}
\end{eqnarray}

\begin{figure}[ht]
\centerline{\epsfxsize=2.2in\epsfbox{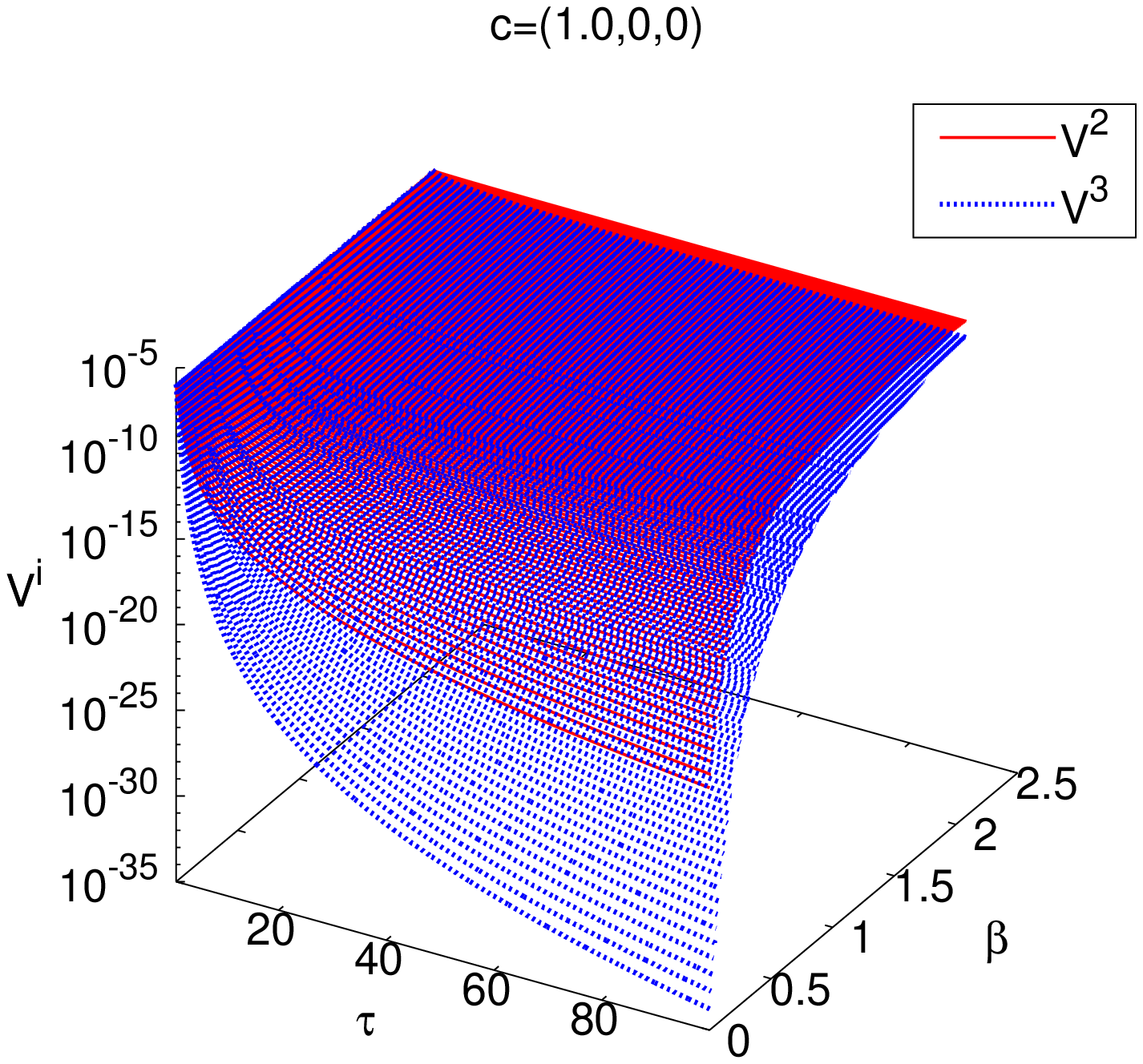}\hspace{0.8cm}
\epsfxsize=2.2in\epsfbox{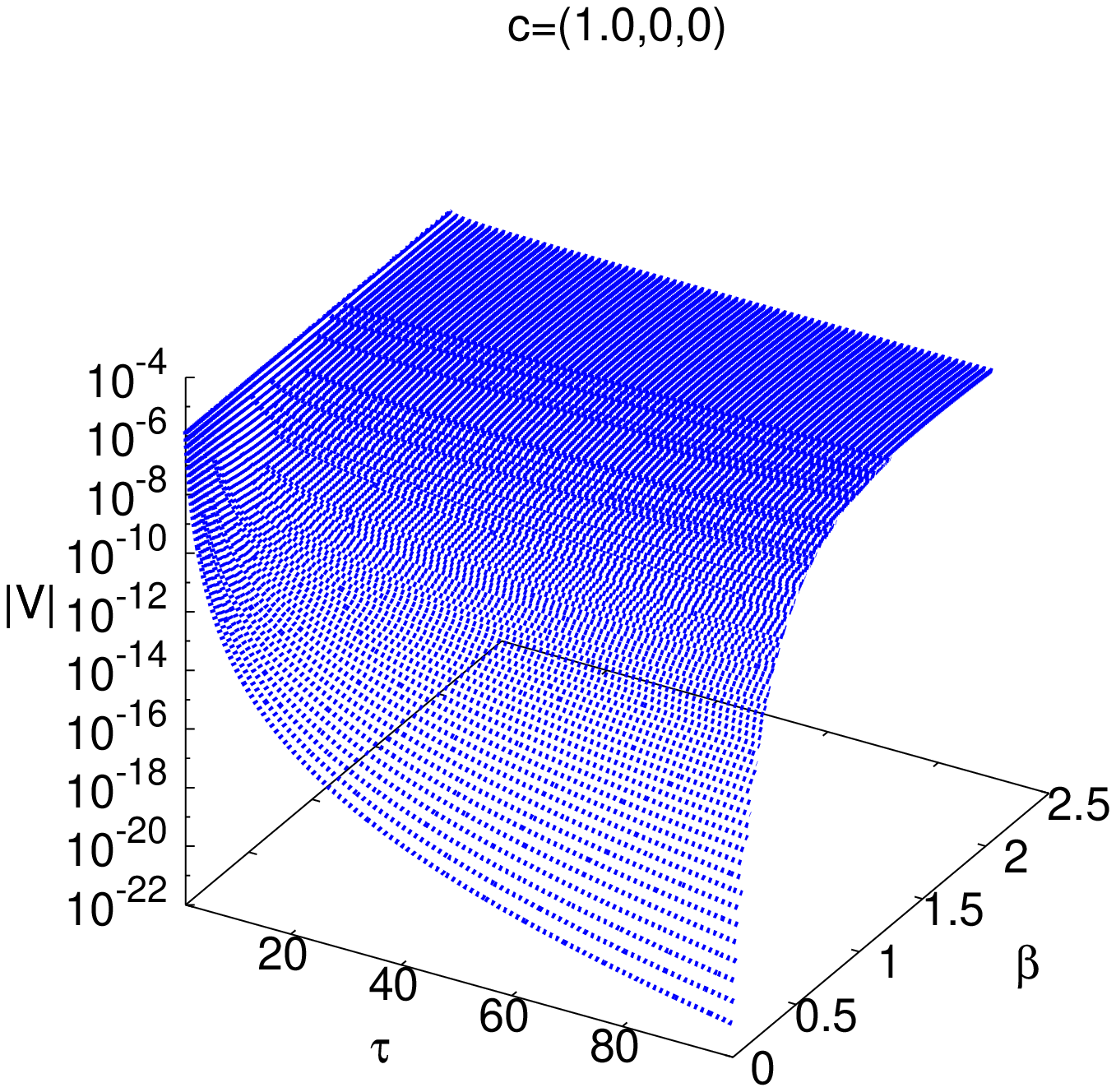}\hspace{0.8cm}
\epsfxsize=2.2in\epsfbox{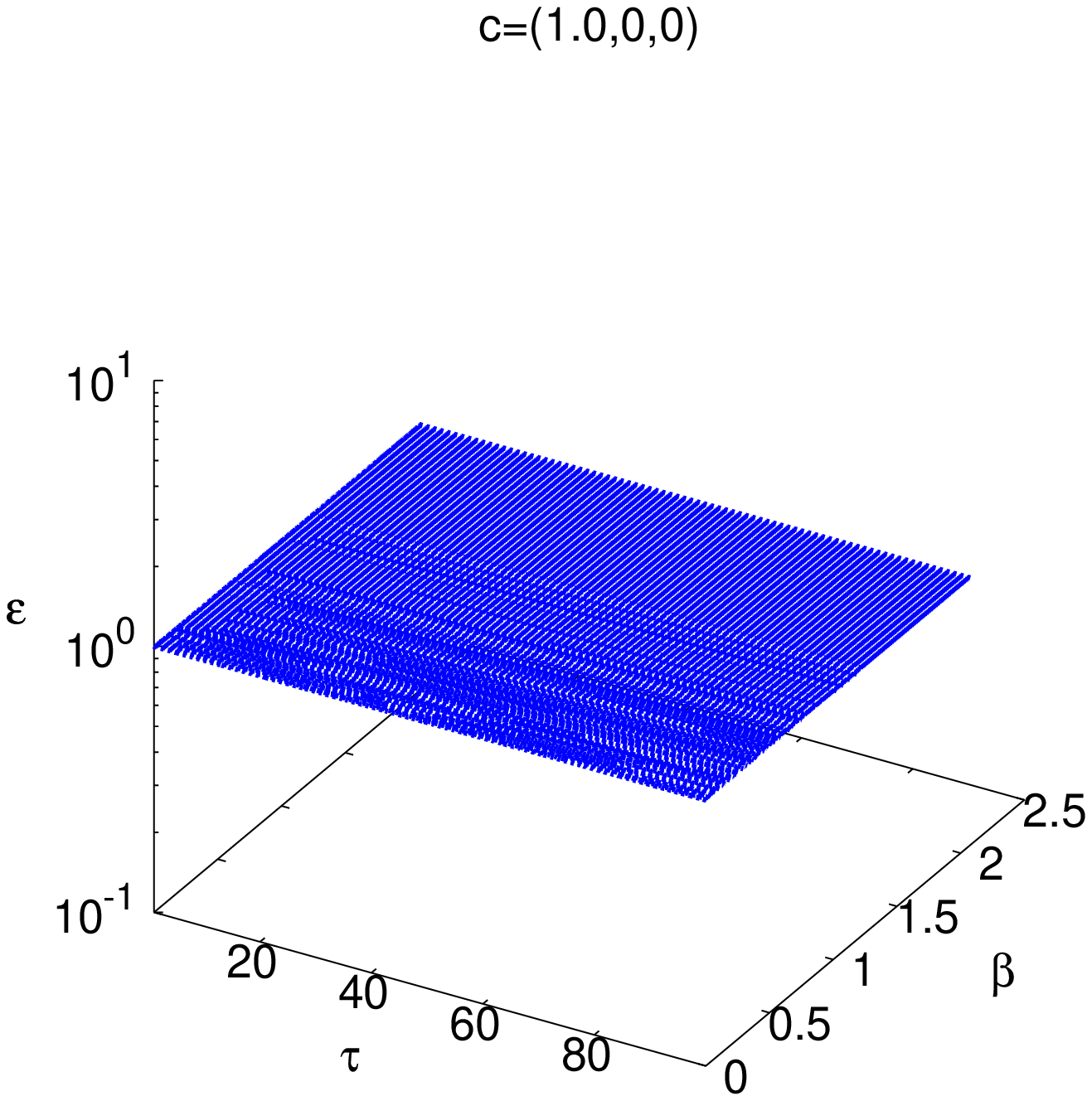}}
\caption{Numerical solutions for the components of the string velocity
  ({\it left}),  $|\vec{V}|=\sqrt{\sum_i \left(\frac{a_i}{a_1}\right)^2(v^i)^2}$ ({\it middle}) and
  $\epsilon$ ({\it right}) for the Kasner solution for the choice
  $c=(1.0,0,0)$.  The initial velocity vector is given by by
  $\vec{V}=(0,V_0,V_0)$ with $V_0=10^{-6}$.}
\label{fig1}
\end{figure}

\begin{figure}[ht]
\centerline{\epsfxsize=2.2in\epsfbox{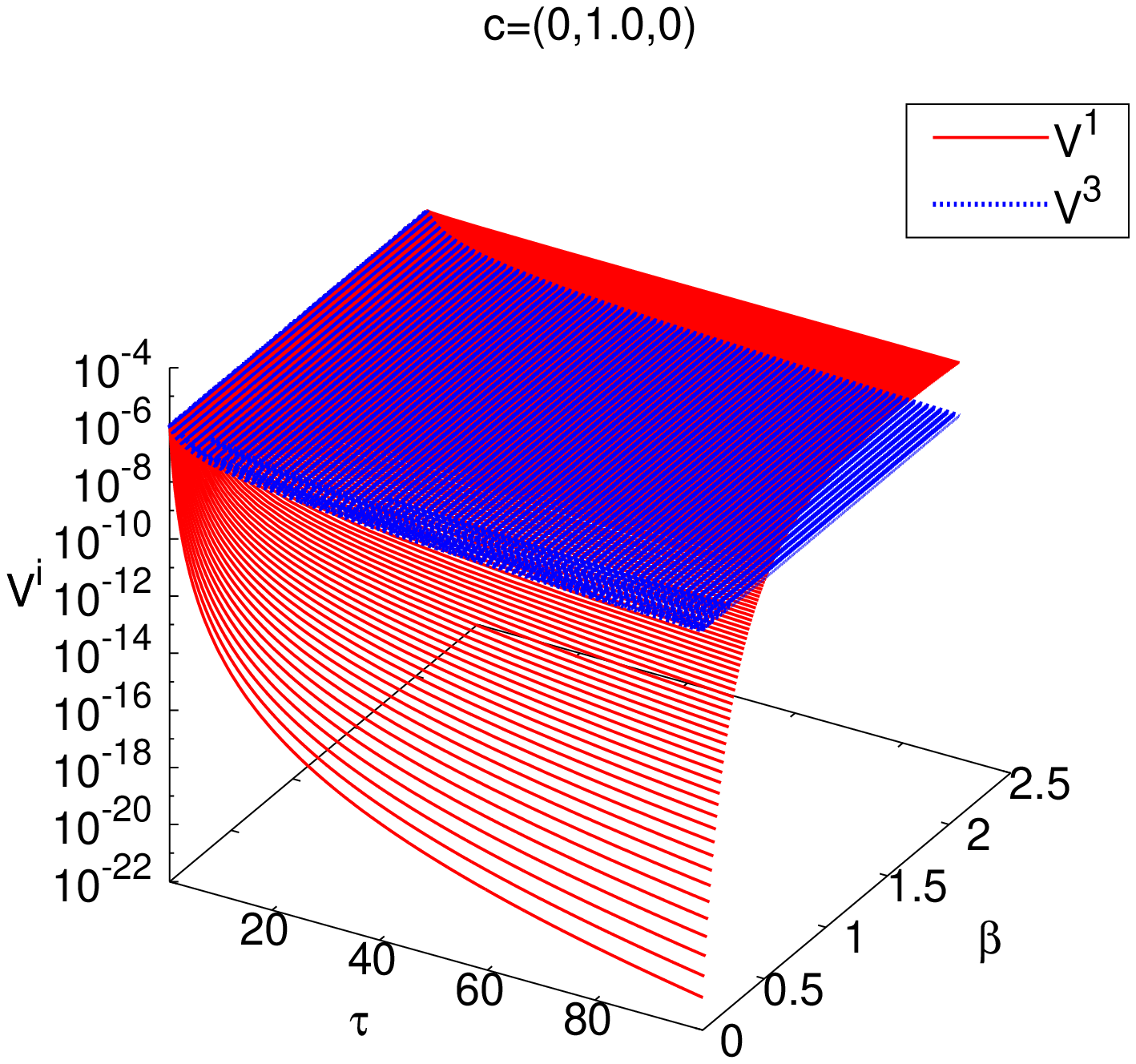}\hspace{0.8cm}
\epsfxsize=2.2in\epsfbox{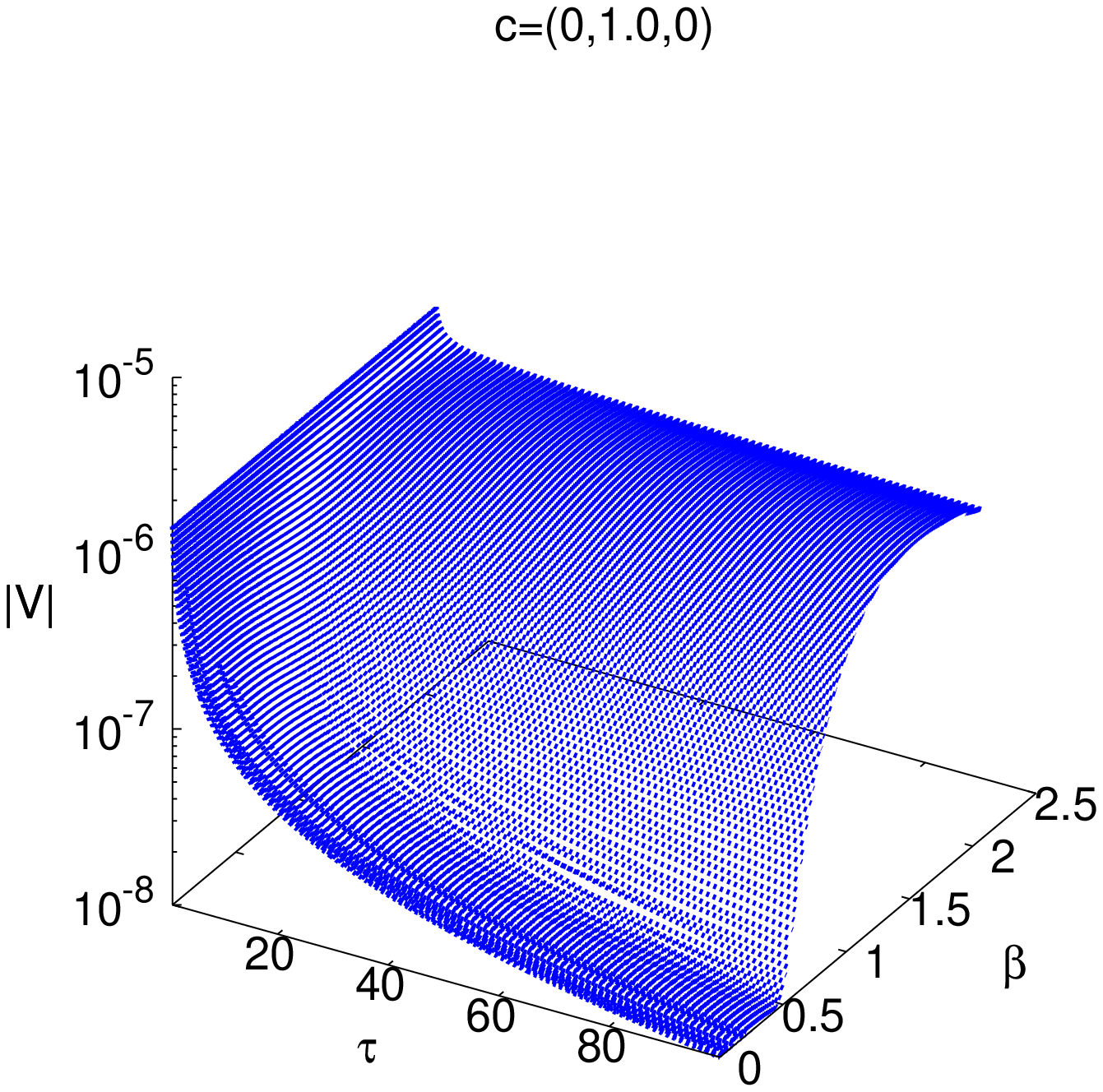}\hspace{0.8cm}
\epsfxsize=2.2in\epsfbox{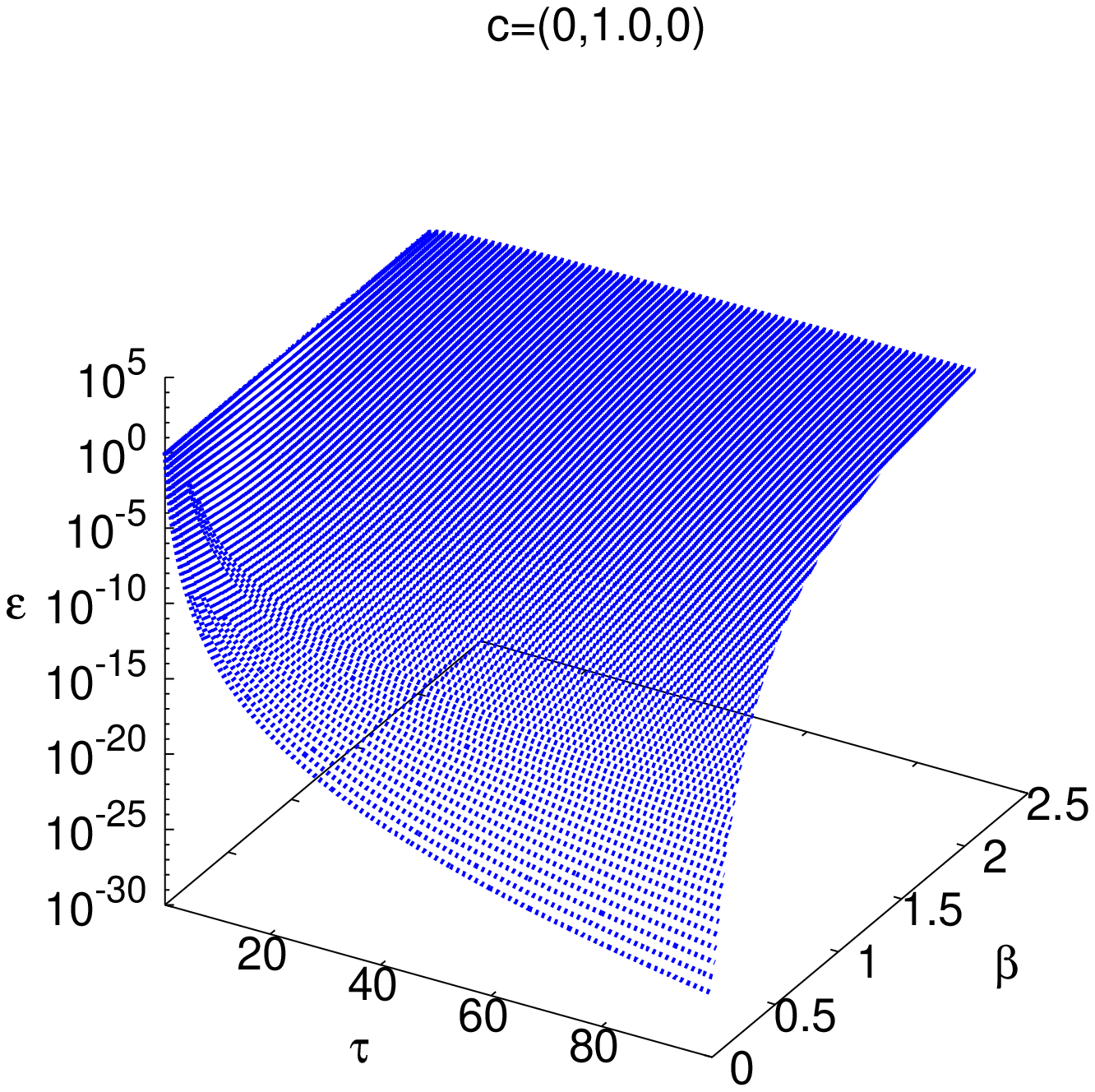}}
\caption{Numerical solutions for the components of the string velocity
  ({\it left}), $|\vec{V}|=\sqrt{\sum_i \left(\frac{a_i}{a_1}\right)^2(v^i)^2}$ ({\it middle}) and
  $\epsilon$ ({\it right}) for the Kasner solution for the choice
  $c=(0,1.0,0)$.  The initial velocity vector is given by
  $\vec{V}=(V_0,0,V_0)$ with $V_0=10^{-6}$.}
\label{fig2}
\end{figure}

In Figs.~\ref{fig1} and \ref{fig2} solutions for anisotropic
backgrounds are presented.  The gauge condition $\dot{x}\cdot
x^{\prime}=\sum_i a_i^2 v^i c^i=0$ is imposed by choosing as initial
conditions  $\vec{v}=(v_0,0,v_0)$ for
$\vec{c}=(0,1,0)$ and $\vec{v}=(0,v_0,v_0)$ for $\vec{c}=(1,0,0)$.
These are two particular configurations chosen in order to illustrate
the general behaviour of a moving straight string in an anisotropic
background.

As can be appreciated in Figs.~\ref{fig1} and \ref{fig2} the
difference in the behaviour of the velocity components is most
significant for small values of $\beta$.  In particular, in the
setting of Fig.~\ref{fig1} the string moves within one collapsing and
one expanding direction. The velocity component in the direction which
is collapsing for $\beta<\pi/2$ completely dominates the movement of
the string.  At $\beta=\pi/2$ this dimension becomes
expanding. However, the corresponding value $p_2<p_3$ (cf.,
Fig.~\ref{fig0}), so that the scale factor is less growing in that
direction.  Moreover, $\epsilon$ is found to be approximately constant
in $\tau$, independently of the value of the Kasner exponents. This
can easily be understood by recalling that in this case the only
non-vanishing component of $c^i$ is $c^1$ and therefore neglecting the
contribution due the velocity components in Eq.~(\ref{eps1}) yields
the approximately constant behaviour of $\epsilon$.

The solution in Fig.~\ref{fig2} picks out one direction which starts
out as an expanding direction and becomes contracting at $\beta=\pi/2$
and one direction which always expands.  The velocity of the string is
dominated by the direction whose scale factor is less growing and
there is a cross over in the magnitudes of the velocity components at
$\pi/6$, where the Kasner exponent $p_3$ becomes larger than the
Kasner exponent $p_1$. So that the scale factor in the latter
direction becomes smaller than in the direction associated with $p_3$.
In this case $\epsilon$ is not a constant, but rather scales
approximately as $\epsilon\sim a_2/a_1$ which is decreasing in $\tau$
for small values of $\beta$ and is slowly growing with time for
$\beta>\pi/2$ (cf., Fig.~\ref{fig0}), with a maximum growth $\propto
\tau^{0.46}$ at $\beta=2\pi/3$.

Therefore, these numerical solutions seem to indicate that the
behaviour of a velocity component is inversely proportional to the
scale factor in that direction. This is similar to the solution in the
isotropic case (cf., Eq.~(\ref{iso})).

%%%%%%%%%%%%%%%%%%%%%%%%%%%%%%%%%%%%%%%%%%%%%%%%%%%%%%%%%%%%%%%%%%%%%%

\section{Perturbations of a straight string}
\label{sec4}

Following Ref.~\cite{vs} to describe small perturbations on a static
straight string we assume
\begin{eqnarray}
x^i=c^i\zeta+\delta x^i~,
\label{xi}
\end{eqnarray}
where $\delta x^i=\delta x^i(\tau,\zeta)$ describes a small
perturbation.  Then Eq.~(\ref{v}) yields after linearization, to first
order in the perturbation,
\begin{eqnarray}
\delta\ddot{x}^i+\left[2\frac{\dot{a}_i}{a_i}
  -\frac{\dot{a}_1}{a_1}+M\right]\delta\dot{x}^i-N(\delta x^i)''=0~,
\label{pert}
\end{eqnarray}
where
\begin{eqnarray}
M\equiv\frac{\sum_j a_j\dot{a}_jc^{j\;\;2}}{\sum_m
  a_m^2(c^m)^2}\hspace{1cm},\hspace{1cm} N=\frac{a_1^2}{\sum_m a_m^2 (c^m)^2}~.
\end{eqnarray}
In an isotropic background this equation reduces to one discussed in \cite{vs} where the condition
$\sum_i\left(c^i\right)^2=1$ is imposed.
The gauge condition Eq.~(\ref{gauge}) leads at linear order to
\begin{eqnarray}
\sum_i a_i^2c^i\delta\dot{x}^i=0~.
\label{constr}
\end{eqnarray}

In order to get a general understanding of the behaviour of the
perturbations in an anisotropic background particular classes of
solutions will be discussed. The simplest way to satisfy the gauge
condition is to assume the following two different classes of
configurations:
\\
\begin{itemize}
\item
{\bf (a)}
one component of $c^i$, call it $c^j$, is non zero and the two 
perturbation components $\delta x^i$, corresponding to $i\neq j$, are non 
vanishing 

\item
{\bf (b)}
one perturbation amplitude $\delta x^i$, call it $\delta x^j$, is non 
vanishing and the two components of $c^i$, corresponding to $i\neq j$, 
are non zero.
\end{itemize}

In the first class are configurations of the form:
\begin{eqnarray}
c^i=(1,0,0) ~~&;&~~ \delta x^i=(0,\delta x^2,\delta
x^3)~,\nonumber\\ c^i=(0,1,0) ~~&;&~~ \delta x^i=(\delta x^1,0, \delta
x^3)~,\nonumber\\ c^i=(0,0,1) ~~&;&~~ \delta x^i=(\delta x^1, \delta
x^2,0)~.
\end{eqnarray}  
The last two cases are basically equivalent; the solutions of one case
can be obtained by interchanging the indices 2 and 3. The first case
is different from the other two due to the fact that the definition of
the conformal time-like variable $\tau$ involves the scale factor
$a_1$.

In the second class are solutions with 
\begin{eqnarray}
c^i=\frac{1}{\sqrt{2}}(1,1,0) ~~;~~ \delta x^i=(0,0, \delta
x^3)~,\nonumber\\ c^i=\frac{1}{\sqrt{2}}(1,0,1) ~~;~~ \delta x^i=(0,
\delta x^2,0)~,\nonumber\\ c^i=\frac{1}{\sqrt{2}}(0,1,1) ~~;~~ \delta
x^i=(\delta x^1,0, 0)~.
\end{eqnarray}  
In both classes the constant vector $c^i$ has been chosen such that
$\sum_i (c^i)^2=1$, as was used in the isotropic case in \cite{vs}.

In the following the perturbation equation, Eq.~(\ref{pert}), will be
solved in both cases assuming
\begin{eqnarray}
\delta x^i=A^i(\tau) e^{ik\zeta}~.
\end{eqnarray}

%%%%%%%%%%%%%%%%%%%%%%%%%%%%%%%%%%%%%%%%%%%%%%%%%%%%%%%%%%%%%%%%%%%%%%

\subsection{Case (a)}
We begin with the case where the string space-time coordinates are
given by
\begin{eqnarray}
\vec{x}=\left(
\begin{array}{c}
\zeta\\
\delta x^2\\
\delta x^3~
\end{array}
\right)~,
\end{eqnarray}
so that the space-like world sheet coordinate $\zeta$ lies along the
space-time $x$-direction. Thus the wavenumber $k$ lies along this
direction as well. So $k$ could be thought of as $k_x$.  The
perturbation equation, Eq.~(\ref{pert}), reads for $i=2,3$
\begin{eqnarray}
\ddot{A}_i+2\frac{\dot{a}_i}{a_i}\dot{A}_i+k^2A^i=0~,
\label{case1}
\end{eqnarray}
which can be solved in terms of Bessel functions \cite{grad} and thus
the perturbation, assuming that it becomes a constant for small
arguments (mimicking the isotropic case \cite{vs}) is given by
\begin{eqnarray}
\delta x^i=(k\tau)^{-\mu_i}J_{\mu_i}(k\tau)e^{ik\zeta}~,
\end{eqnarray}
where $J_{\mu_i}(z)$ is a Bessel function of the first kind and
$\mu_i\equiv (2\alpha_i-1)/2$, where the $\alpha_i$ are the exponents
in the scale factors (cf., Eq.~(\ref{scale})).  The physical
wavelength of the wave traveling in $x$-direction is given by
$\lambda_{\rm phys}=2\pi a_1(\tau)/k$.

For $k\tau\ll 1$ the amplitudes $\delta x^i$ in the $y$- and
$z$-directions approach constant values. Since these are the
coordinate values, the physical amplitudes are determined by $\delta
x^i_{\rm phys}\equiv a_i(\tau)\delta x^i$, ($i=2,3$).  Therefore the
ratio $\delta x^i_{\rm phys}/\lambda_{\rm phys}\sim a_i/a_1$. Thus
contrary to the isotropic case the string is not conformally
stretched, but rather becomes wigglier or straightens out, which means
that the amplitude grows or diminishes with respect to the wavelength,
respectively, depending on the ratio of the two scale factors.

In the opposite limit $k\tau\gg 1$ the physical amplitudes $\delta
x^i_{\rm phys}$ become constant. Therefore, since the physical wavelength
grows as $a_1$, the string becomes straighter, which is the same
behaviour as found in the isotropic case \cite{vs}. If, however, the
$x$-direction is a collapsing direction the string becomes wigglier
with time which is the case for $\beta>\pi/2$ (cf., Fig.~\ref{fig1}).

Choosing a different configuration in this class
the string will be  described by 
\begin{eqnarray}
\vec{x}=\left(
\begin{array}{c}
\delta x^1\\
\zeta\\
\delta x^3
\end{array}
\right)~.
\end{eqnarray}
So that the space-like world-sheet coordinate $\zeta$ lies along the
space-time $y$-direction. Thus the wavenumber $k$ lies along this
direction as well. So $k$ could be thought of as $k_y$.  The
perturbation equation, Eq.~(\ref{pert}), reads for $i=1,3$
\begin{eqnarray}
\ddot{A}^i+\left[2\frac{\dot{a}_i}{a_i}-\frac{\dot{a}_1}{a_1}
  +\frac{\dot{a}_2}{a_2}\right]\dot{A}^i+\left(\frac{a_1}{a_2}\right)^2k^2A^i=0~,
\label{case2}
\end{eqnarray}
which can be solved in terms of Bessel functions \cite{grad} and
yields the perturbations
\begin{eqnarray}
\delta x^i=(k\tau)^{\mu_i}J_{\nu_i}\left[\beta(k\tau)^{\gamma}\right]e^{ik\zeta},
\end{eqnarray}
where 
\begin{eqnarray}
\mu_1&=&\frac{1-\alpha_1-\alpha_2}{2}~, \nonumber\\ 
% \hspace{4cm}
\mu_3&=&\frac{1-2\alpha_3+\alpha_1-\alpha_2}{2}
\nonumber\\ \nu_1&=&-\frac{1-\alpha_1-\alpha_2}{2(1+\alpha_1-\alpha_2)}~,\nonumber\\ 
%\hspace{3.2cm}
\nu_3&=&-\frac{1-2\alpha_3+\alpha_1-\alpha_2}{2(1+\alpha_1-\alpha_2)}~,
\nonumber\\ \beta&=&\frac{a_{1,*}}{a_{2,*}}\frac{(k\tau_*)^{\alpha_2-\alpha_1}}
{|1+\alpha_1-\alpha_2|}~,
\nonumber\\ \gamma&=&1+\alpha_1-\alpha_2~.
\label{pert2}
\end{eqnarray}
As can be seen from Fig.~\ref{fig0} the parameter $\gamma=(1-p_2)/(1-p_1)$ is always
positive in the case of the Kasner solutions considered here.
Since $\gamma>0$ the comoving perturbation amplitudes $\delta x^i$,
for $i=1, 3$, approach a constant in the limit $k\tau\ll 1$. However,
the ratio of the physical amplitude over the physical wavelength
becomes in this case $\delta x^i_{\rm phys}/\lambda_{\rm phys}\sim a_i/a_2$.
Therefore, it depends on the ratio of the scale factors if the string
is straightened or becomes wigglier.  In the limit $k\tau\gg 1$ the
physical amplitude becomes a constant. Therefore, the ratio of the
physical perturbation amplitude over the physical wavelength goes as
$1/a_2$. So that in the case that this is an expanding direction the
string straightens and in the opposite case, that is for a collapsing
direction it becomes wigglier.

\begin{figure}[ht]
\centerline{\epsfxsize=3.4in\epsfbox{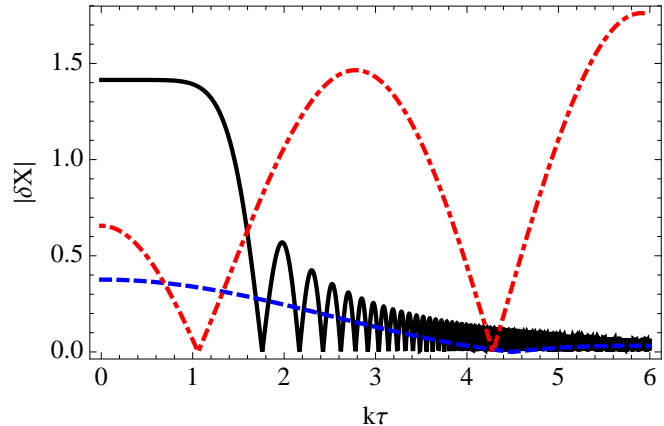}}
\caption{The total co-moving perturbation amplitude $\delta
  x\equiv\sqrt{(\delta x^1)^2+(\delta x^3)^2}$ for the Kasner solution
  with $(p_1,p_2,p_3)=(2/3,-1/3,2/3)$ {\sl
    (black, solid line)}, the isotropic solution
  $p=p_1=p_2=-1/3$ {\sl (red, dot-dashed line)} and
  $p=p_1=p_2=2/3$ {\sl (blue, dashed line)}.  }
\label{fig5}
\end{figure}

In Fig.~\ref{fig5} the total co-moving perturbation amplitude $\delta
x\equiv\sqrt{|\delta x^1|^2+|\delta x^3|^2}$ is shown for
$(p_1,p_2,p_3)=(2/3,-1/3,2/3)$ which corresponds to $\beta=\pi/6$.  It
is compared with the corresponding total perturbation amplitude in the
isotropic backgrounds with $p=p_1=p_2=-1/3$ and $p=p_1=p_2=2/3$, where
$p$ is the exponent of the scale factor using cosmic time.

%%%%%%%%%%%%%%%%%%%%%%%%%%%%%%%%%%%%%%%%%%%%%%%%%%%%%%%%%%%%%%%%%%%%%%

\subsection{Case (b)}

It is assumed that the  string is described by
\begin{eqnarray}
\vec{x}=\left(
\begin{array}{c}
\zeta/\sqrt{2}\\
\zeta/\sqrt{2}\\
\delta x^3
\end{array}
\right)~.
\end{eqnarray}
So that the space coordinates $x$ and $y$ satisfy $x=y$, which implies
that the string lies along this diagonal in the ($x$-$y$)-plane. Moreover, 
$k=\sqrt{k_x^2+k_y^2}$.  The perturbation equation reads
\begin{eqnarray}
\ddot{A}^3+\frac{1}{\tau}\left[2\alpha_3-\alpha_1
  +\frac{\alpha_1+\alpha_2\left(\frac{a_{2,*}}{a_{1,*}}
    \right)^2\left(\frac{\tau}{\tau_*}\right)^{2(\alpha_2-\alpha_1)}}{
    1+\left(\frac{a_{2,*}}{a_{1,*}}\right)^2
    \left(\frac{\tau}{\tau_*}\right)^{2(\alpha_2-\alpha_1)}}\right]
\dot{A}^3+\frac{2k^2}{1+\left(\frac{a_{2,*}}
  {a_{1,*}}\right)^2\left(\frac{\tau}{\tau_*}\right)^{2(\alpha_2-\alpha_1)}}A^3=0~.
\end{eqnarray}
In this case it seems more difficult to find an exact solution, so we
will consider different regimes.  Assuming $\tau_*$ is some initial
time then $\tau/\tau_*\geq 1$. Now assuming that $\alpha_2-\alpha_1>0$
then the third term in the coefficient of the $\dot{A}^3$ reduces to
$\alpha_2$ and the time-dependent term dominates in the denominator of
the coefficient of $A^3$. Thus the resulting equation corresponds to
Eq.~(\ref{case2}). In the case where $\alpha_2-\alpha_1<0$ an equation
similar (with an additional factor of $\sqrt{2}$ in the argument of
the Bessel function) to Eq.~(\ref{case1}) results. 

Therefore the behaviour of the ratio between the physical amplitude
and the physical wavelength is similar to the one discussed in case
(a). The physical wavenumber in case (b) is given by
$$k^2_{\rm
  phys}=\left(\frac{k_x}{a_1}\right)^2+\left(\frac{k_y}{a_2}\right)^2~$$
implying that the physical wavelength reads
\begin{equation}
\lambda_{\rm phys}\sim \frac{1}{k_{\rm phys}}
=\frac{a_2}{\left[\left(\frac{a_2}{a_1}\right)^2k_x^2+k_y^2\right]
  ^{\frac{1}{2}}}~.
\end{equation}
Thus, for $a_2/a_1\gg 1$ the physical wavelength $\lambda_{\rm
  phys}\sim a_1$, which implies that for $k\tau\ll 1$ the ratio of the
physical perturbation over the physical wavelength, and thus the
wiggliness of the string, evolves as $a_3/a_1$. In the opposite case
$k\tau\gg 1$ this ratio evolves as $1/a_1$.  Moreover, in the case for
$a_2/a_1\ll 1$ the physical wavelength evolves as $\lambda_{\rm
  phys}\sim a_2$.  The resulting behaviour of the ratio of the
physical amplitude over physical wavelength is then described as in
the previous discussion with $a_1$ replaced by $a_2$.

%%%%%%%%%%%%%%%%%%%%%%%%%%%%%%%%%%%%%%%%%%%%%%%%%%%%%%%%%%%%%%%%%%%%%%%
\subsection{Discussion}

We have found that in an anisotropic background it is no longer true
that irregularities on a straight string are frozen-in on super-horizon
scales as it is the case in isotropic backgrounds where the shape of a
string remains unchanged.  In the case of an anisotropic background,
however, the wiggliness of the string can augment or diminish
depending on the time evolution of the ratio of the two scale factors
involved.

This might mean for example that strings from an anisotropic phase
could have much more small scale structure at the time of re-entry
into the horizon during the standard radiation or matter dominated
phases. This would imply the generation of more gravitational
radiation \cite{Sakellariadou:1990ne} which might be used to put
limits on the global anisotropy of the early universe, which could
imply constraints on the minimum duration of a inflationary era.

The behaviour on sub-horizon scales is basically the same as in the
isotropic case, that is the wiggliness scales with $1/a_i$. Here, due
the nature of the Kasner solution, instead of decaying wigglinesss
might increase if the universe is contracting in that particular
direction.

%%%%%%%%%%%%%%%%%%%%%%%%%%%%%%%%%%%%%%%%%%%%%%%%%%%%%%%%%%%%%%
\section{String loops}
\label{sec5}

There are many different types of string loop solutions beginning with
the simplest case which in an isotropic background is a circular loop
lying in one of the coordinate planes. Most explicit solutions have
been found in flat space-time, such as the two-parameter solutions
\cite{KT,T} and generalizations thereof (see for example
Ref.~\cite{vs}). However, loops have also been considered in curved
backgrounds such as Friedmann-Lema\^itre-Roberston-Walker
\cite{thomps,li} or Kantowski-Sachs, Bianchi I and IX {\cite{LD}.
  Loops in an axisymmetric background have been studied numerically in
  Ref.~\cite{am2}.
  In this section the scale factors are assumed to be general and not 
  necessarily the Kasner solutions used in the previous sections.

We will begin by considering periodic solutions in general. The gauge
condition (Eq.~(\ref{gauge})) and $\epsilon$ (cf., Eq.~(\ref{eps}))
take the form
\begin{eqnarray}
\sum_ia_i^2\dot{x}^ix^{i\;'}=0~,\\ 
\sum_i\left(\frac{a_i}{a_1}\right)^2\left[\dot{x}^{i\;
    2}+\epsilon^{-2}(x^{i\; '})^2\right]=1~,
\end{eqnarray} 
respectively.  These two equations together with Eqs.~(\ref{e1}) and
(\ref{v}) form the set of equations determining the dynamics of a
cosmic string in an anisotropic background.  The equations reduce in
the case of a flat space-time to the known ones (cf., e.g.,
Ref.~\cite{vs}).  In flat space-time Eq.~(\ref{v}) reduces to a wave
equation which is solved in general by (e.g., Ref.~\cite{vs})
\begin{eqnarray}
\vec{x}(\zeta,\tau)=\frac{1}{2}\left[\vec{b}_+(\zeta_+)
  +\vec{b}_{-}(\zeta_{-})\right]~,
\label{xb}
\end{eqnarray}
where $\zeta_{\pm}\equiv\zeta\pm\tau$ and $b_{\pm}$ are arbitrary
functions subject to the constraint
$\partial_+\vec{b}_+^{\, 2}=\partial_{-}\vec{b}_{-}^{\, 2}=1$ where
$\partial_{\pm}\equiv \partial/\partial\zeta_{\pm}$.  Since we
are interested in periodic solutions we use the form of
Eq.~(\ref{xb}) which yields to
\begin{eqnarray}
\sum_i\left(\frac{a_i}{a_1}\right)^2\left(\partial_{+}b^i_{+}\right)^2=
\sum_i\left(\frac{a_i}{a_1}\right)^2\left(\partial_{-}b^i_{-}\right)^2~,\label{c1}\\ 
\frac{1}{2}\left(1+\epsilon^{-2}\right)\sum_i\left(\frac{a_i}{a_1}\right)^2
\left(\partial_{-}b^i_{-}\right)^2
+\frac{1}{2}\left(\epsilon^{-2}-1\right)\sum_i\left(\frac{a_i}{a_1}\right)^2\partial_+b_+^i
\partial_-b_-^i=1~,
\label{c2}\\
\dot{\epsilon}=-\epsilon\left[\frac{\dot{a}_1}{a_1}+\frac{1}{4}
\left(1-\epsilon^{-2}\right)\sum_i\frac{a_i\dot{a}_i}{a_1^2}\left[(\partial_+b_-^i)^2
+(\partial_+b_+ ^i)^2\right]-\frac{1}{2}\left(1+\epsilon^{-2}\right)
\sum_i\frac{a_i\dot{a}_i}{a_1^2}\partial_-b_-\partial_+b_+\right]~,\label{c3}\\
\left(1-\epsilon^{-2}\right)\left[\partial^2_{--}b_-^i+\partial^2_{++}b_+^i\right]=
\left[2\frac{\dot{a}_i}{a_i}
+\frac{\dot{\epsilon}}{\epsilon}
-\frac{\epsilon'}{\epsilon^3}\right]\partial_-b_-^i
-\left[2\frac{\dot{a}_i}{a_i}+\frac{\dot{\epsilon}}{\epsilon}
+\frac{\epsilon'}{\epsilon^3}\right]\partial_+b_+^i~.\label{c4}
\end{eqnarray}

So far no approximations have been made and these equations are exact.
Clearly the flat space-time solution can be recovered.  In flat
space-time loop solutions are described by
$\vec{b}_{\pm}(\zeta+L)=\vec{b}_{\pm}(\zeta)$ which leads to
$\vec{x}(\zeta+L/2,\tau+L/2)=\vec{x}(\zeta,\tau)$, where $L$ is the
invariant length of the loop \cite{vs}.  Moreover, if the time-scale
of oscillation is comparable to the loop length, in other words if the
relation $\vec{x}(\zeta+L/2,\tau+L/2)=\vec{x}(\zeta,\tau)$ is
satisfied, one may deduce that the motion of the loop is relativistic.
From Eqs.~(\ref{c1})-({\ref{c4}) it can be seen that the effect of the
  expansion of the universe, which is not a periodic function of time,
  is to inhibit the periodicity of the solutions.  However, in the
  regime where $L$ is smaller than the Hubble time, which is
  equivalent to considering string loops whose sizes are smaller than
  the horizon \footnote{In numerical simulations with periodic
    boundary conditions, all strings are closed; sub-horizon strings
    are thus often referred to as loops and super-horizon ones as
    infinite strings.}, the departure from periodicity of
the solutions is small. Therefore considering sub-horizon solutions
and time-scales where the evolution of the scale factors can be
neglected, it is a good approximation to consider $\epsilon\simeq 1$ a
constant. In this case the last equation is identically satisfied and
the functions $b_{\pm}^i$ are arbitrary subject to the constraint
given by Eqs.~(\ref{c1}) and (\ref{c2}).  Note that a similar approach
has been considered in Ref.~\cite{thomps} for isotropic backgrounds.

In the anisotropic case the constraints imply
\begin{eqnarray}
\sum_i\left(\frac{a_i}{a_1}\right)
^2(\partial_-b_-^i)^2=\sum_i\left(\frac{a_i}{a_1}\right)
^2(\partial_+b_+^i)^2=1.
\label{const}
\end{eqnarray}
Therefore, in the case of an anisotropic space-time periodic solutions
are described by vector functions $\partial_{-}\vec{b}_-$ and
$\partial_{+}\vec{b}_+$ which are closed functions on an ellipsoid,
described in general by the equation $x^2/A^2+y^2/B^2+z^2/C^2=1$ where
$A$, $B$ and $C$ are the equatorial radii and the polar radius,
respectively. These radii are given by the inverse of the ratio of the
scale factors in the different directions and the scale factor
$a_1$. In the case of an isotropic background these solutions are on a
unit sphere as it is the case in a flat space-time \cite{vs}.
Considering the constraint equations, Eq.~(\ref{const}), formally the
flat space-times solutions are recovered by assuming the relation
\begin{eqnarray}
b_{\pm}^{i\;\;({\rm flat})}=\frac{a_i}{a_1}\;b_{\pm}^{i\;\;({\rm aniso})}~,
\label{transf}
\end{eqnarray}
which holds on time-scales when the expansion of the universe can be
neglected.  This means that well inside the horizon the condition for
intersection of two closed curves determined by $\partial_+\vec{b}_+$
and $-\partial_ -\vec{b}_-$ is determined by the same condition as in
flat space-time, that is $\partial_+\vec{b}_+^{\;({\rm
    flat})}=-\partial_-\vec{b}_-^{\;({\rm flat})}$. Therefore, it is
expected that the number of cusps in this limit is the same as in the
case of a flat background. 

Calculating the square of the proper velocity of the string at such an
intersection it is found that, using Eq. (\ref{const}), 
\begin{eqnarray}
\vec{V}^2=\sum_i\left(\frac{a_i}{a_1}\right)^2\left(\partial_+
b^i_+\right)^2=1,
\end{eqnarray}
so that these are luminal points exactly as  in an isotropic background.

Moreover, as in flat space-time the behaviour of the string at the
cusp can be studied by expanding around $\zeta_{\pm}=0$ \cite{dv}
(cf., also Ref.~\cite{vs}, \cite{bo}).  Thus choosing $\zeta_{\pm}$ such that the
cusp is at $\zeta_{\pm}^{({\rm c})}=0$ as well as
$\vec{x}(\zeta_{\pm}^{({\rm c})})=0$, truncating the expansion at third
order, one gets
\begin{eqnarray}
\vec{b}_{\pm}(\zeta_{\pm})=\vec{n}^{({\rm c})}\zeta_{\pm}
+\frac{1}{2}\vec{b}^{\;(2)}_{\pm}\zeta_{\pm}^2
+\frac{1}{6}\vec{b}^{\;(3)}_{\pm}\zeta_{\pm}^3~,
\end{eqnarray}
where $\vec{n}^{({\rm c})}\equiv\partial_+\vec{b}_+=-\partial_-\vec{b}_-$
and the notation $\vec{b}_{\pm}^{\;(i)}$ indicates the $i$-th
derivative with respect to the argument.  Moreover, $\vec{n}^{({\rm c})}$
and $\vec{b}^{(i)}$ are calculated at the cusp.  Differentiating the
constraint Eq.~(\ref{const}) and using that at the cusp $V^i=\frac{a_i}{a_1}n^{(c)\; i}$ 
results in
\begin{eqnarray}
\sum_i V^i \,\frac{a_i}{a_1}x^{(2)i}&=&0~,
\label{cu1}\\
\sum_iV^i \,\frac{a_i}{a_1}x^{(3)i}&
=&\frac{1}{2}\sum_i\left(\frac{a_i}{a_1}\right)^2\left[\left(b_{-}^{(2) i}\right)^2
-\left(b_+^{(2) i}\right)^2\right].
\label{cu2}
\end{eqnarray}
Equation (\ref{cu1}) implies that at the cusp two orthogonal directions
can be defined using relation Eq.~(\ref{transf}). This corresponds to
locally defining an orthonormal frame.  Therefore, expressing the
approximate solution at the cusp in terms of the flat space solution
at $\tau=0$ yields to
\begin{eqnarray}
x^i=\frac{1}{2}\frac{a_1}{a_i}x^{({\rm flat})(2)
  i}(0)\zeta^2+\frac{1}{6}\frac{a_1}{a_i}x^{({\rm flat})(3) i}(0)\zeta^3~,
\end{eqnarray}
where $x^{({\rm flat})(2) i}$ (and $x^{({\rm flat})(3) i}$) denote
the second (and third, respectively) derivative of the $i$-th
component of the  three-vector $x$ written in terms of left- and
right-movers.  Assuming that $\vec{x}^{({\rm flat})(2)}(0)$ and
$\vec{x}^{({\rm flat})(3)}(0)$ are orthogonal directions, say
$\hat{e}^{(2)}=\hat{y}$ and $\hat{e}^{(1)}=\hat{x}$ then in the flat
space-time or isotropic case the set of parametric equations is given
by $x=(1/6)\zeta^3$ and $y=(1/2)\zeta^2$ which yields the standard
behaviour $y\propto x^{2/3}$ \cite{vs,and}. However, in the
anisotropic case the parametric equations include the ratio of the
scale factors. Choosing the directions as before, one finds
$x=(1/6)\zeta^3$ and $y=(a_1/2/a_2)\zeta^2$.
\begin{figure}[ht]
\centerline{\epsfxsize=4.1in\epsfbox{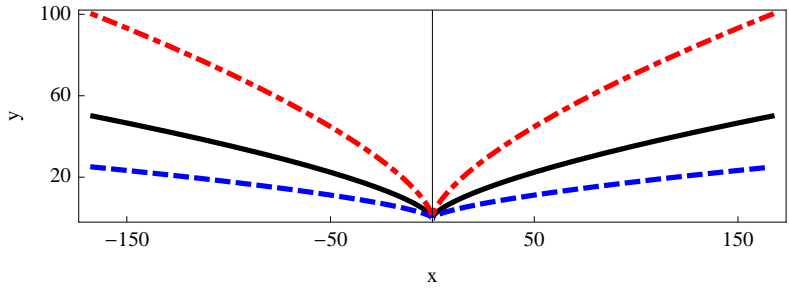}}
\caption{Shape of a cusp in the flat space-time ({\it black, solid
    line}), in an anisotropic space-time with $a_1=0.5 a_2$ ({\it red,
    dot-dashed line}) and for $a_1=2a_2$ ({\it blue, dashed line}).
  $\vec{x}^{({\rm flat})(3)}(0)$ lies in the direction $x$ and
  $\vec{x}^{({\rm flat})(2)}(0)$ lies in the direction $y$.}
\label{fig6}
\end{figure}
Thus as can be seen in Fig.~\ref{fig6} the shape, namely the opening
of the cusp is changed in the anisotropic case in comparison with the
flat or isotropic case. Moreover, as it is expected the shape depends
on the orientation of the cusp, that is in which direction it is
assumed that the second and third derivatives lie at the cusp.

As a particular example the generalization of the elliptical loop
solution in flat space-time \cite{vv} will be considered.  Assuming
that the string is already well inside the horizon, in the anisotropic
background this solution takes the form in terms of the variables
$\zeta_{\rm}$:
\begin{eqnarray}
\vec{x}=\frac{1}{2}\left(
\begin{array}{c}
\sin\zeta_{+}+\sin\zeta_-\\
\frac{a_1}{a_2}\left(\cos\zeta_+ +\cos\zeta_-\right)\cos\frac{\phi}{2}\\
\frac{a_1}{a_3}\left(\cos\zeta_- -\cos\zeta_+\right)\sin\frac{\phi}{2}
\label{ellip}
\end{array}
\right)~,
\end{eqnarray}
where $\phi$ determines the shape of the ellipse.  Whereas in an
isotropic background $\phi=0$ and $\phi=\pi$ correspond to an
oscillating circular loop that collapses to a point and a rotating
double line, this is no longer the case in an anisotropic background.

We have thus shown that an anisotropic background may leave its
fingerprints on the dynamics of cosmic string loops.  In an isotropic
geometry cusps are defined as points where the string moves
momentarily at the speed of light. 
It was shown that this also holds in  anisotropic backgrounds.  
The physical velocity of the string reaches the speed of light at a cusp.
Thus cusps are important since they
create strong gravitational pulses which propagate at the speed of
light along the beam and they are extending from the cusp in the
direction perpendicular to the string velocity. 
If the loops entering the horizon have a richer small-scale structure and less pronounced cusps,
then one should expect a fan-like pattern of gravity wave beams, like
the one emitted by kinks, rather the single beam structure
characteristic of radiation emitted by cusps. Moreover, the difference
in the shape of the cusps should leave an imprint on the effect of
gravitational pulses from cusps on particle trajectories.

\section{Conclusions}

Different configurations of cosmic strings have been considered in an
anisotropic background.  The motivation for considering such an
anisotropic case is that the phase before the universe reached
isotropy during inflation has left an imprint on a cosmic string
network. This has been suggested in Ref.~\cite{am2} where the equation
for a string loop has been solved numerically in an axisymmetric
background.

A static straight string is a solution of an anisotropic
background. Perturbations of a straight string depend on the global
anisotropy of the background.  In particular, it was found that the
wiggliness of a string on super-horizon scales depends on the ratio of
two scale factors. Therefore, once the string re-enters the horizon
during the standard radiation or matter domination it might be much
more or much less wigglier than in the isotropic case changing the
emitted gravitational radiation. Thus opening the possibility by using
gravitational wave constraints to put limits on the global anisotropy
in the very early universe.

Studying a static circular loop, it was found that the anisotropic
evolution of the universe tends to impede periodic solutions.  Writing
the equations of motion and the constraint equations in terms of left-
and right-moving solutions it was found that if the size of the string
loop is smaller than the horizon or equivalently if the expansion of
the universe can be neglected periodic solutions exist.

Moreover, it was found that the derivatives of the right- and
left-moving vectors describe closed curves on an ellipsoid rather than
the unit sphere as it is the case in an isotropic universe and flat
space-time.  Furthermore, the condition for the formation of cusps is
the same in flat space-time, so that the number of cusps does not
change.  In an anisotropic space-time the formation of cusps is
determined by the intersection of two-dimensional curves on an
ellipsoid.  However, with respect to the flat space-time the shape or
opening of the cusp does change in an anisotropic space-time 
\footnote{This is different to the case of cusps in models with
higher dimensions where the numbers as well as the shape of cusps
change \cite{calletal}. In fact for $n$ extra dimensions the
probability of cusp formation is zero since instead of intersections
of a one dimensional object on a 2-sphere (in the isotropic case) one
has to consider intersections on a $(2+n)$-sphere and thus
intersections are unlikely \cite{cg}.  This led \cite{calletal} to
consider near cusp events which are rounded cusps.}.
Finally, the proper string velocity at the cusp reaches the speed of light just like in 
an isotropic background.
Thus one might expect the gravity waves radiation from
cusps are of the same magnitude as in an isotropic background whereas the
importance of gravity waves from the small-scale structure of the
strings as they enter the horizon may turn out to be more important.

If the size of the loop is of the size of the horizon or even larger
in general the solutions are expected to be non periodic and so the
number of loops should be very small.  However, the distribution of
shapes of loops formed during the anisotropic phase has to be left to
a full numerical simulation. This also applies to the general shape of the strings, since 
one would not expect the strings to be described by  simple random walks
and the network characterized by just one characteristic length
if a cosmic string network forms in an anisotropic background \cite{vs,as}.

Finally, the conical string geometry also has observational implications. However,
the deficit angle in an axisymmetric Bianchi I model is the same as in 
an isotropic background. Thus one may conclude that for a sub-horizon string
observational effects, such as the formation of double images of light sources located behind the
string position, are similar to those of a string in a Minkowski background.

%%%%%%%%%%%%%%%%%%%%%%%%%%%%%%%%%%%%%%%%%%%%%%%%%%%%%%%%%%%%%%%%%%%%%
\vspace{1cm}
\noindent
{\bf Acknowledgments}\\ It is a pleasure to thank Larissa Lorenz for
discussions. 
K.E.K. gratefully acknowledges financial support from 
Spanish Science Ministry grants FPA2009-10612, FIS2009-07238 and CSD2007-00042.
The work of M.S. is partially supported by the Sciences
\& Technology Facilities Council (STFC--UK), Particle Physics
Division, under the grant ST/G000476/1 ``Branes, Strings and Defects
in Cosmology''.

%%%%%%%%%%%%%%%%%%%%%%%%%%%%%%%%%%%%%%%%%%%%%%%%%%%%%%%%%%%%%%%%%%%%

\end{document}